\documentclass[
    aps,
    pra,
    reprint,
    superscriptaddress
]{revtex4-2}

\usepackage{amsmath}
\usepackage{amssymb}
\usepackage{bm}

\usepackage{graphicx}
\usepackage{dcolumn}
\usepackage{booktabs}
\usepackage{subcaption}
\usepackage{overpic}

\usepackage{xcolor}
\usepackage[
    colorlinks=true,
    citecolor=blue,
    linkcolor=blue,
    urlcolor=blue
]{hyperref}

\begin{document}

\title{Electron Correlation Enables Phase-Coherent One-Attosecond Pulse Trains}

\author{Andrés Marchisio}
\email{marchisio.andres@gtiit.edu.cn}
\affiliation{Department of Physics, Guangdong Technion - Israel Institute of Technology,
241 Daxue Road, Shantou, Guangdong, China, 515063}
\affiliation{Technion – Israel Institute of Technology, Haifa, 32000, Israel}

\author{Isobel McSweeney}
\affiliation{ICFO-Institut de Ciencies Fotoniques, The Barcelona Institute of Science and Technology, Av. Carl Friedrich Gauss 3, 08860 Castelldefels (Barcelona), Spain}



\author{Paraskevas Tzallas}
\affiliation{Foundation for Research and Technology-Hellas, Institute of Electronic Structure \& Laser, GR-70013 Heraklion (Crete), Greece}
\affiliation{Center for Quantum Science and Technologies (FORTH-QuTech), GR-70013 Heraklion (Crete), Greece}
\affiliation{ELI-ALPS, ELI-Hu Non-Profit Ltd., Dugonics tér 13, H-6720 Szeged, Hungary}

\author{Maciej Lewenstein}
\affiliation{ICFO-Institut de Ciencies Fotoniques, The Barcelona Institute of Science and Technology, Av. Carl Friedrich Gauss 3, 08860 Castelldefels (Barcelona), Spain}

\author{Marcelo F. Ciappina}
\email{marcelo.ciappina@gtiit.edu.cn}
\affiliation{Department of Physics, Guangdong Technion - Israel Institute of Technology,
241 Daxue Road, Shantou, Guangdong, China, 515063}
\affiliation{Technion – Israel Institute of Technology, Haifa, 32000, Israel}
\affiliation{Guangdong Provincial Key Laboratory of Materials and Technologies for Energy Conversion,
Guangdong Technion – Israel Institute of Technology, 241 Daxue Road, Shantou, Guangdong, China, 515063}

\date{\today}

\begin{abstract}
Attosecond synthesis is ultimately a phase problem: a broad spectrum produces an ultrashort waveform only if its harmonics remain phase-locked. We show theoretically that correlated two-electron high-harmonic generation in helium driven by a long, multicycle laser pulse supports a train of soft-x-ray bursts with durations approaching 1 as. Using a two-electron strong-field approximation, we calculate the complex harmonic spectrum and reconstruct the temporal emission while retaining its full intrinsic spectral phase, rather than imposing a flat-phase approximation. Despite the trajectory-dependent phase accumulated by two continuum electrons, the correlation-extended plateau contains a broad phase-coherent region reaching the keV range. Its superposition produces reproducible bursts separated by one half-cycle of the driving field. These results identify electron correlation not only as a mechanism for extending the high-harmonic cutoff, but also as a route toward phase-coherent x-ray waveforms on the zeptosecond timescale.

\end{abstract}

\maketitle

High-order harmonic generation (HHG) provides a direct route from intense
optical fields to coherent extreme-ultraviolet and x-ray radiation and forms
the basis of attosecond science~\cite{Paul2001,Krausz2009}. For a long,
multicycle driving field, the phase-locked odd harmonics constitute a frequency
comb whose temporal counterpart is an attosecond pulse train, with consecutive
bursts separated by one half of the optical period. The duration and temporal
structure of each burst, however, are determined not by the spectral bandwidth
alone but by the harmonic spectral phase. Its first derivative gives the
frequency-dependent emission time, whereas its curvature describes the
intrinsic attosecond chirp generated by the underlying electron
trajectories~\cite{Mairesse2003,LopezMartens2005}. The experimental frontier has
reached pulse durations of a few tens of attoseconds and photon energies
extending through the water window and into the keV
regime~\cite{Gaumnitz2017,Cousin2017,Popmintchev2012}. Approaching the
$1~\mathrm{as}$ scale requires a qualitatively more demanding combination: an
exceptionally broad spectrum that remains mutually coherent across hundreds of
harmonic orders.

Within the conventional single-active-electron picture, the available
bandwidth is restricted by the cutoff energy
\begin{equation}
    E_{\mathrm{cut}} = I_p + 3.17U_p,
\end{equation}
where $I_p$ is the ionization potential and $U_p$ is the ponderomotive
energy ($U_p=E_0^2/4\omega_0^2$, $E_0$ being the laser electric field peak amplitude and $\omega_0$ its carrier frequency)~\cite{Lewenstein1994}. Increasing the driving wavelength can move the
cutoff toward the keV range, but at the cost of severe wave-packet spreading,
a rapidly decreasing single-atom yield, and increasingly demanding macroscopic
phase matching~\cite{Popmintchev2012}. Correlated two-electron dynamics offer a
fundamentally different route. In nonsequential double recombination, two
electrons ionized at different field extrema propagate in the continuum and
recombine simultaneously, releasing their combined kinetic and binding
energies as a single high-energy photon~\cite{Koval2007,Hansen2016,Chacon2016}.
This process generates secondary plateaus with cutoff scalings substantially beyond the conventional single-electron limit.
Recent experiments in UV-driven helium have revealed a
correlation-induced secondary plateau extending beyond the conventional
single-electron cutoff to approximately $280~\mathrm{eV}$, using a
$400~\mathrm{nm}$ driving field with a peak intensity of about
$2\times10^{15}~\mathrm{W/cm^2}$~\cite{Wang2026}. This spectral extension cannot be explained
by the standard single-active-electron cutoff law and has been attributed to
the simultaneous recombination of two correlated electrons \cite{McSweeney2026}. The observation provides compelling experimental evidence for
double-electron recombination and demonstrates that electron correlation can
open a coherent x-ray emission channel far beyond the conventional
single-electron HHG cutoff. Two correlated trajectory families are possible,
depending on whether the electrons are ionized approximately half a cycle or
one full cycle apart before recombining simultaneously, leading to
\begin{equation}
\begin{aligned}
E_{\mathrm{cut}}^{(\pi)}
&= I_p^{(1)}+I_p^{(2)}+4.7U_p,\\
E_{\mathrm{cut}}^{(2\pi)}
&= I_p^{(1)}+I_p^{(2)}+5.5U_p.
\end{aligned}
\end{equation}
Here, $I_p^{(1)}$ and $I_p^{(2)}$ denote the first and second
sequential ionization potentials, respectively. For helium, for instance,
$I_p^{(1)}\simeq24.6~\mathrm{eV}$ corresponds to
$\mathrm{He}\rightarrow\mathrm{He}^{+}+e^{-}$, whereas
$I_p^{(2)}\simeq54.4~\mathrm{eV}$ corresponds to
$\mathrm{He}^{+}\rightarrow\mathrm{He}^{2+}+e^{-}$. Their sum,
$I_p^{(1)}+I_p^{(2)}\simeq79.0~\mathrm{eV}$, is the total
double-ionization threshold recovered when both electrons recombine
synchronously with the parent ion and emit a single photon.

A cutoff extension, however, does not by itself imply the generation of shorter
pulses. Temporal reconstructions based on a flat spectral phase provide only a
transform-limited estimate and do not establish whether the extended harmonics
can interfere constructively in time. This distinction becomes especially
important for double-electron emission: the harmonic phase contains the actions
accumulated by two electronic wave packets, their distinct ionization times,
their continuum excursion times, and the interference between competing return
trajectories. Whether the resulting broad near-keV spectrum preserves
sufficient phase synchronization to generate one-attosecond waveforms therefore
remains an open question.

Here we address this question by calculating the complex two-electron HHG
dipole response generated in helium by a long, multicycle driving pulse. We extract the
spectral phase throughout the
correlation-induced plateau and reconstruct the emitted radiation directly from
the complex harmonic amplitudes. We find that a broad region of the extended
plateau remains sufficiently phase locked to produce a regular train of bursts
with durations approaching $1~\mathrm{as}$, repeated every half-cycle of the
driving field. The result survives the inclusion of the intrinsic dipole
phase and is therefore not a consequence of an assumed flat phase. Electron
correlation thus acts simultaneously as an energy-upconversion mechanism and
as a coherent temporal-synthesis channel, opening a route toward x-ray pulse
trains at the zeptosecond frontier.


Following the two-electron saddle-point framework developed by McSweeney \textit{et al.}~\cite{McSweeney2026}, the high-harmonic phase is dictated by the stationary action associated with the double ionization and joint recombination dynamics,
\begin{equation}
\begin{aligned}
    \phi(t_r, t_{1i}, t_{2i}, p_1, p_2) = &-\frac{1}{\hbar} S_{1e}(p_1, t_{1i}, t_{2i}) \\ 
    &- \frac{1}{\hbar}S_{2e}(p_1, p_2, t_{2i}, t_r) + q\omega_0 t_r,
\end{aligned}
\label{eq:phase_def}
\end{equation}
where $t_{1i}$ and $t_{2i}$ represent the first and second ionization times, $t_r$ is the joint recombination time, $p_1, p_2$ are the canonical momenta of the two correlated electrons, and $q$ is the harmonic order.
The single- and two-electron semiclassical actions read:
\begin{align}
S_{1e} &= \int_{t_{1i}}^{t_{2i}} d\bar{t} \left[ \frac{\left(p_1 - \frac{e}{c}A(\bar{t})\right)^2}{2m} + I_p^{(1)} \right], \label{eq:S1e} \\
S_{2e} &= \int_{t_{2i}}^{t_r} d\bar{t} \left[ \frac{\left(p_1 - \frac{e}{c}A(\bar{t})\right)^2}{2m} + \frac{\left(p_2 - \frac{e}{c}A(\bar{t})\right)^2}{2m} +I_p^{(1)}+I_p^{(2)} \right]. \label{eq:S2e}
\end{align}

\begin{figure}[t]
\centering

  \begin{subfigure}[b]{0.9\linewidth}
    \centering
    \begin{overpic}[width=\linewidth]{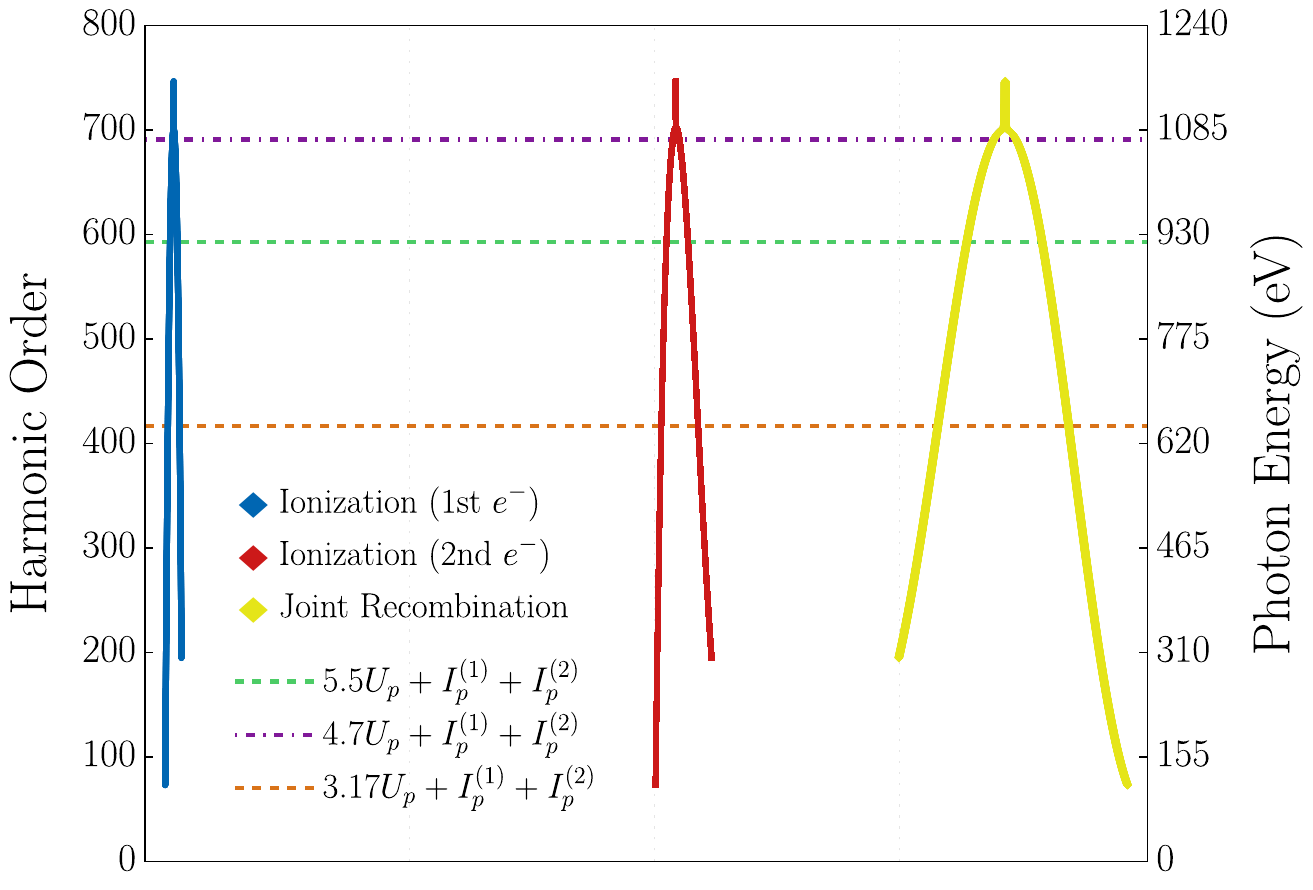}

      \put(15, 61){\textbf{(a)}}
    \end{overpic}
    \phantomcaption 
    \label{fig:trajectories_a}
  \end{subfigure}

  \vspace{-0.3cm} 
  \begin{subfigure}[b]{0.9\linewidth}
    \centering
    \begin{overpic}[width=\linewidth]{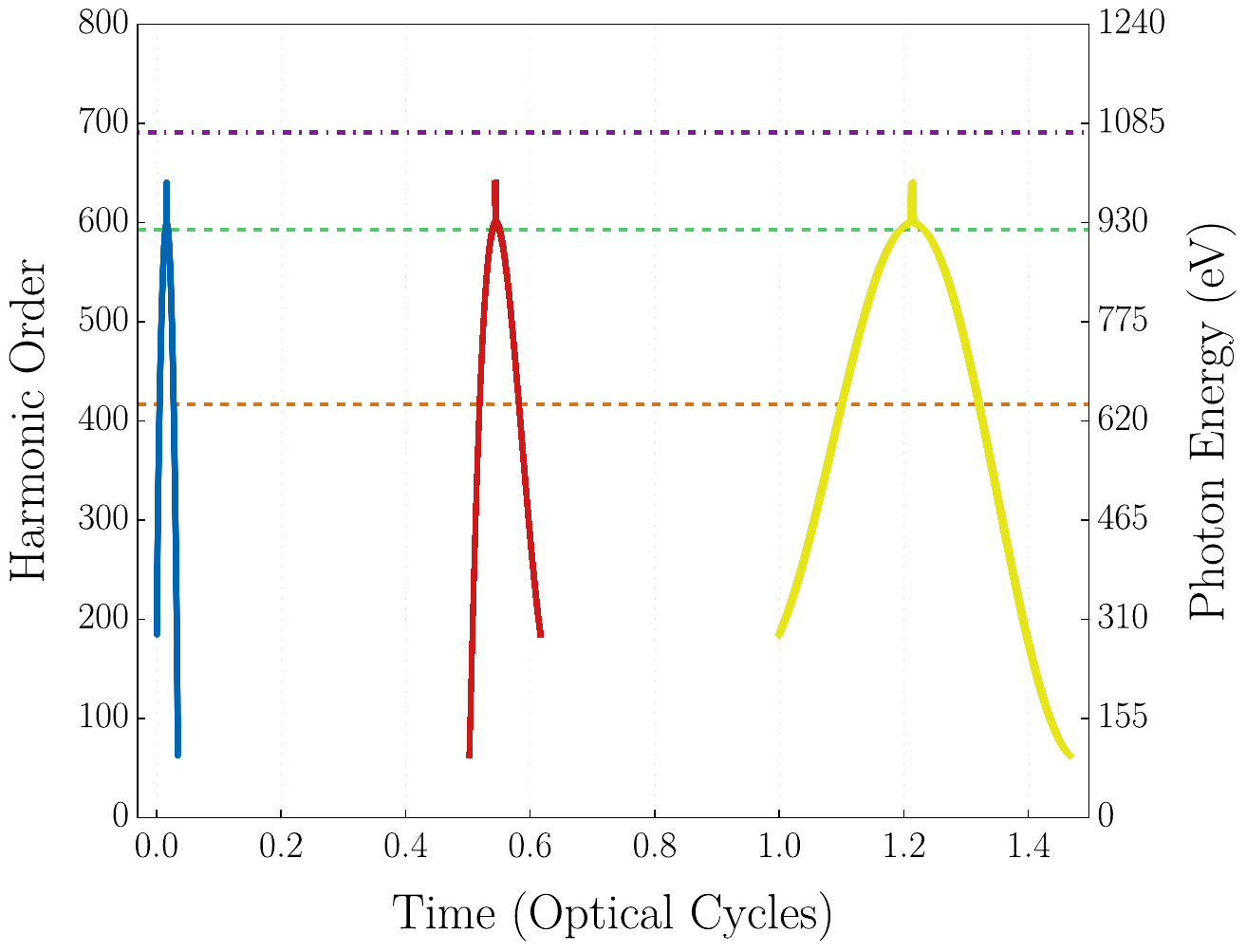}
      \put(15, 70){\textbf{(b)}}
    \end{overpic}
    \phantomcaption
    \label{fig:trajectories_b}
  \end{subfigure}

  \caption{\textbf{Quantum orbit trajectory dynamics in two-electron HHG.} Real parts of ionization times ($t_{1i}$, $t_{2i}$) and joint recombination time $t_r$ as a function of harmonic order. (a)~$t_{2i} - t_{1i} \approx 2\pi/\omega_0$ (one optical cycle delay), exhibiting a maximum energy cutoff near $5.5\,U_p + I_p^{(1)}+I_p^{(2)}$, and (b)~$t_{2i} - t_{1i} \approx \pi/\omega_0$ (half optical cycle delay), extending the harmonic plateau up to the $4.7\,U_p + I_p^{(1)}+I_p^{(2)}$ cutoff energy.}
  \label{fig:trajectories}
\end{figure}

Extrema of Eq.~(\ref{eq:phase_def}) yield the saddle-point equations determining the complex quantum trajectories \cite{pisanty2016attosecond,pisanty2020rbsfa,Nayak2019}. Importantly, these saddle-point calculations are tailored exclusively to the joint double recombination process, enabling us to isolate and selectively examine the distinct ionization delay channels corresponding to $\Delta t=~t_{2i}~-~t_{1i} \approx \pi/\omega_0$ and $\Delta t \approx 2\pi/\omega_0$. In a realistic physical scenario, both delay channels coexist simultaneously alongside single-electron recombination pathways \cite{Wang2026}. As shown in Fig.~\ref{fig:trajectories}, the separation between electron ionization events $\Delta t $ critically dictates the kinetic energy gained in the laser field. For a full-cycle delay ($\Delta t \approx 2\pi/\omega_0$), the cutoff extends to $5.55\,U_p + I_p^{(1)}+I_p^{(2)}$ [Fig.~\ref{fig:trajectories}(a)], whereas a half-cycle delay ($\Delta t \approx \pi/\omega_0$) reduces the harmonic cutoff up to $4.7\,U_p + I_p^{(1)}+I_p^{(2)}$ [Fig.~\ref{fig:trajectories}(b)].

Within this framework, the quantum orbits can be classified into two distinct classes based on the excursion times of the individual electrons. While the first electron consistently follows a long trajectory prior to the second ionization event, the second electron can be emitted into either a short or a long continuum path before recombination. Consequently, we designate the combined two-electron process as a \textit{short} or \textit{long} quantum trajectory depending strictly on the continuum dynamics of the second electron.

Averaging the kinetic energy terms over optical cycles yields the linear scaling of the dipole phase $\phi$ with driving intensity $I$:
\begin{equation}
\phi_1(I, \Omega) \approx \alpha_{2e}(\Omega) U_{p}(I) + \phi_0(\Omega),
\label{eq:linear_phase}
\end{equation}
where $\Omega = q \omega_{0}$ and $U_{p}(I)=\frac{I}{4\omega_0^2}$ reflects the accumulated ponderomotive action in the continuum. By evaluating the trajectory times of flight, the dipole phase coefficient is explicitly given by
\begin{equation}
\alpha_{2e}(\Omega) = \left[ (t_{2i} - t_{1i}) + 2(t_r - t_{2i}) \right],
\label{eq:alpha2e_explicit}
\end{equation}
combining the single-electron continuum drift period $(t_{2i} - t_{1i})$ with twice the excursion period $(t_r - t_{2i})$ during the two-electron propagation stage.

\begin{figure}[ht]
\centering
  \begin{subfigure}[b]{0.48\linewidth}
    \centering
    \begin{overpic}[width=\linewidth]{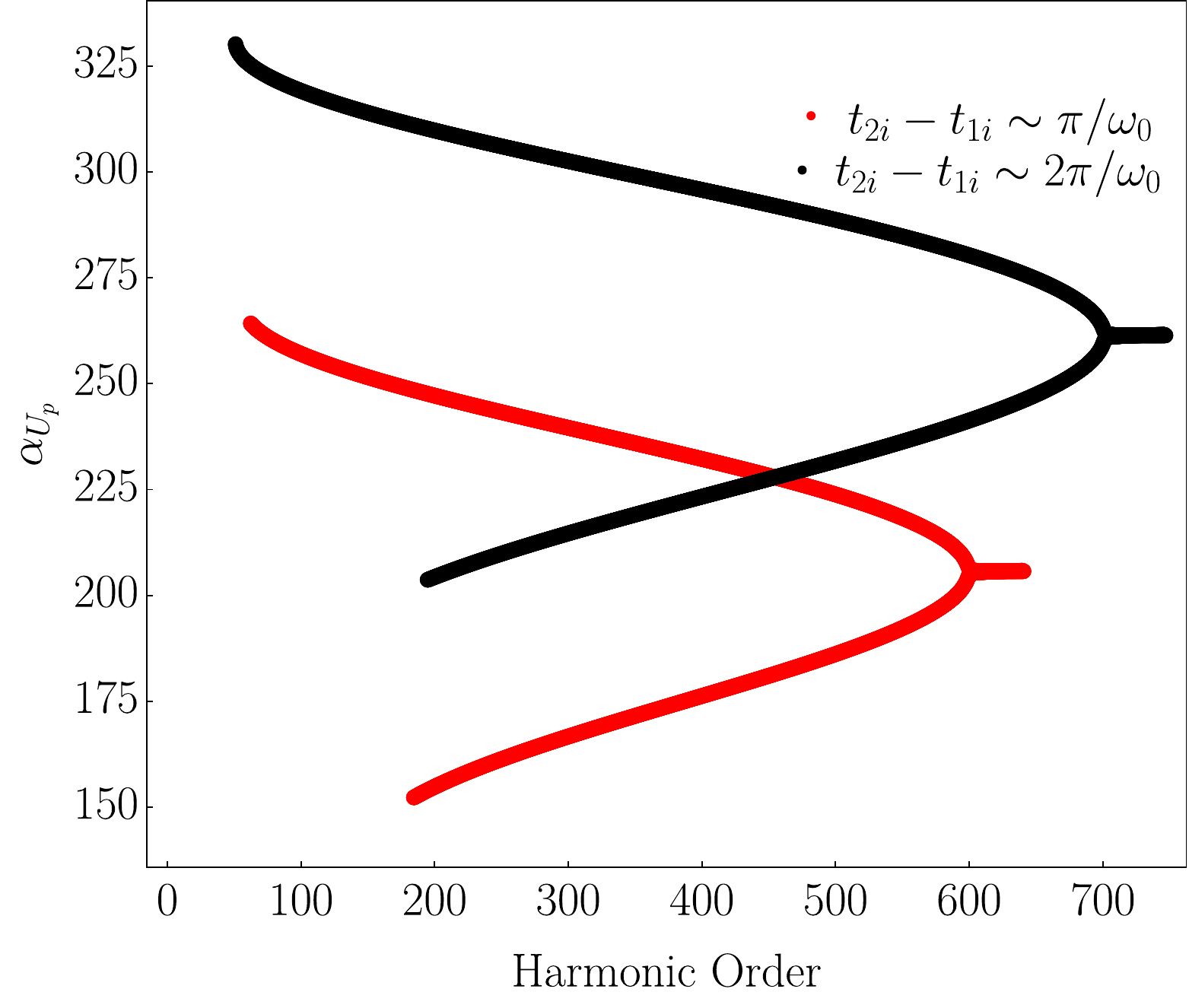}
      \put(15, 68){\textbf{(a)}}
    \end{overpic}
    \phantomcaption
   \label{fig:alpha_vs_omega}
  \end{subfigure}
  \hfill
  \begin{subfigure}[b]{0.50\linewidth}
    \centering
    \begin{overpic}[width=\linewidth]{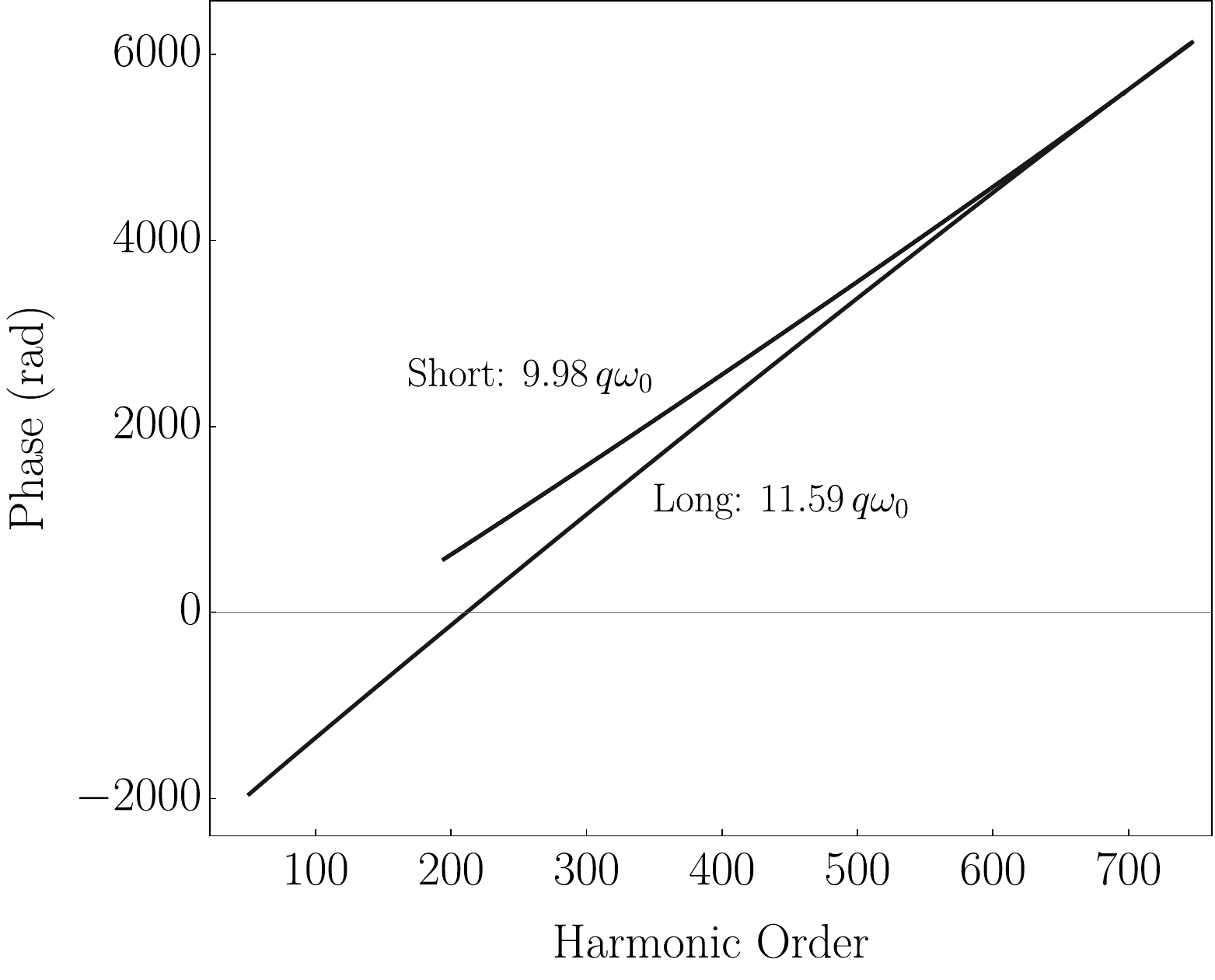}
      \put(18, 68){\textbf{(b)}}
    \end{overpic}
    \phantomcaption
    \label{fig:phi_vs_omega}
  \end{subfigure}

  \vspace{-0.3cm} 

  \begin{subfigure}[b]{0.48\linewidth}
    \centering
    \begin{overpic}[width=\linewidth]{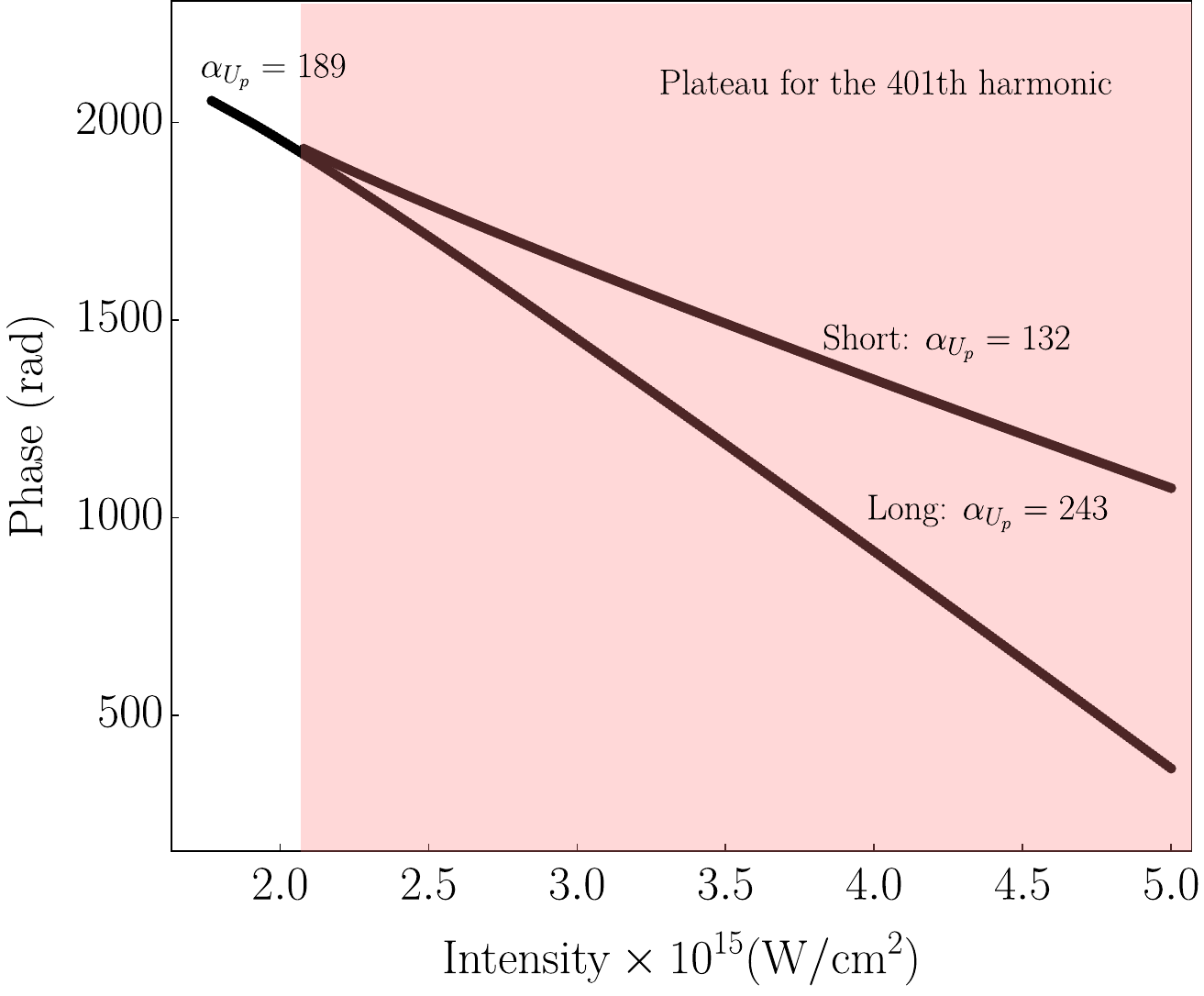}
      \put(15, 60){\textbf{(c)}}
    \end{overpic}
    \phantomcaption
    \label{fig:phase_intensity_47}
  \end{subfigure}
  \hfill
  \begin{subfigure}[b]{0.48\linewidth}
    \centering
    \begin{overpic}[width=\linewidth]{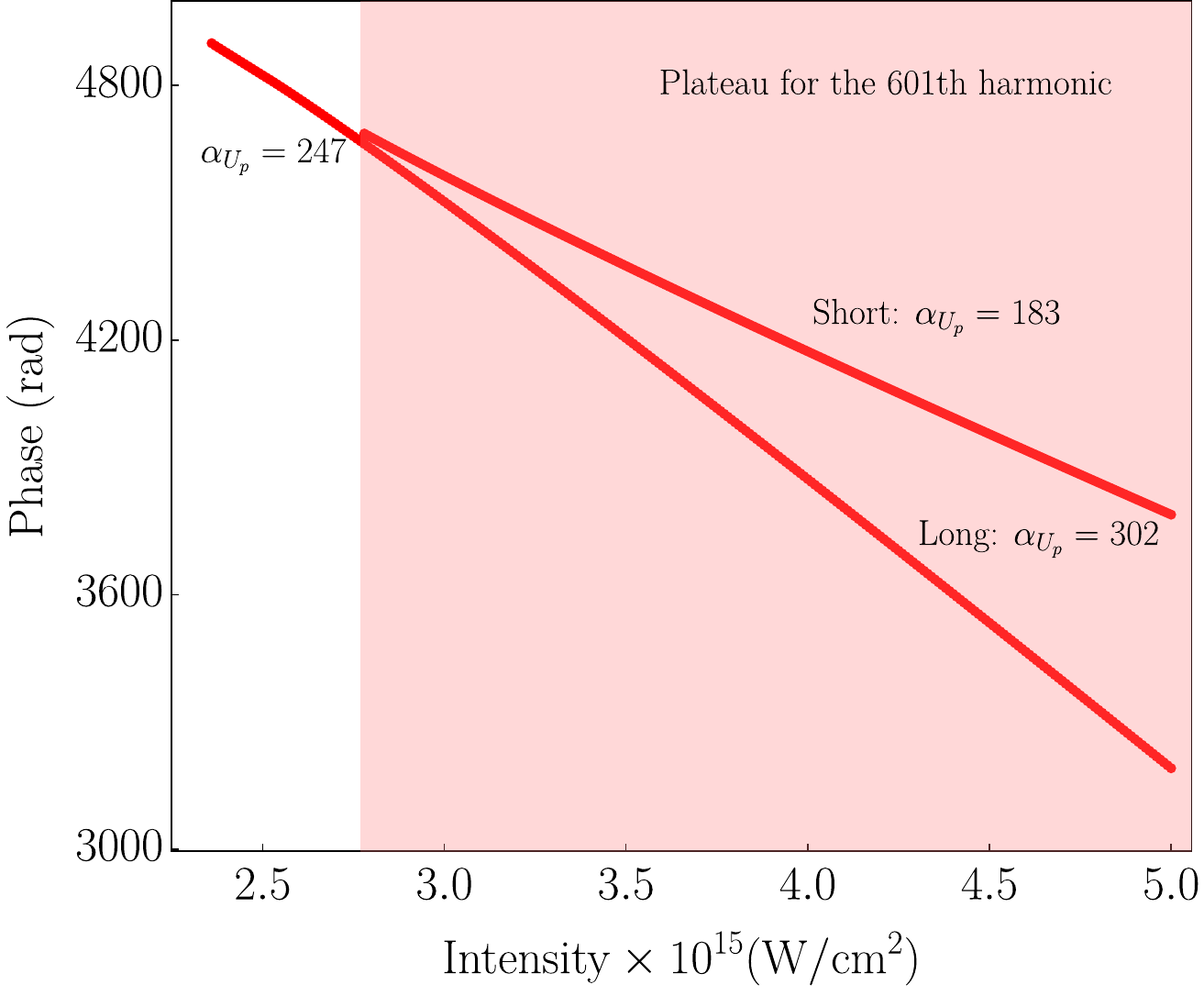}
      \put(18, 60){\textbf{(d)}}
    \end{overpic}
    \phantomcaption
    \label{fig:phase_intensity_55}
  \end{subfigure}

  \caption{\textbf{Dipole phase and $\alpha_{2e}$ scaling characteristics.} (a)~Phase slope $\alpha_{2e}$ as a function of harmonic order for one-cycle ($\Delta t \approx 2\pi/\omega_0$) and half-cycle ($\Delta t \approx \pi/\omega_0$) ionization delays. (b)~Total dipole phase $\phi(\Omega)$ across the plateau for one-cycle ($\Delta t \approx 2\pi/\omega_0$) ionization delay (see Eq.~\eqref{eq:phase_def}), for both, short and long branches. (c)~Dipole phase $\phi(I)$ versus laser intensity for the 401st harmonic under half-cycle ionization delay ($\Delta t \approx \pi/\omega_0$), showing short ($\alpha_{U_p}=132$) and long ($\alpha_{U_p}=243$) trajectory branches. (d)~Dipole phase $\phi(I)$ for the 601st harmonic under full-cycle ionization delay ($\Delta t \approx 2\pi/\omega_0$), with corresponding short ($\alpha_{U_p}=183$) and long ($\alpha_{U_p}=302$) branch slopes.}
  \label{fig:phase_scaling}
\end{figure}

The phase properties of the two-electron emission are systematically analyzed in Fig.~\ref{fig:phase_scaling}. Fig.~\ref{fig:phase_scaling}(a) illustrates the spectral variation of $\alpha_{2e}(\Omega)$, which represents the proportionality constant in a simplified trajectory model, and demonstrates distinct slope regimes governed by the ionization delay $\Delta t$. Likewise, Fig.~\ref{fig:phase_scaling}(b) shows the complete harmonic phase profile $\phi(\Omega)$ calculated using the full stationary action phase from Eq.~\eqref{eq:phase_def} for an ionization delay of $\Delta t \approx 2\pi/\omega_0$, explicitly resolving both short and long quantum paths. In Figs.~\ref{fig:phase_scaling}(c) and \ref{fig:phase_scaling}(d), we track the intensity-dependent phase $\phi(I)$ separated into short and long trajectories for a delay of $\Delta t \approx \pi/\omega_0$ (harmonic $q=401$) and $\Delta t \approx 2\pi/\omega_0$ (harmonic $q=601$), respectively. As expected, both panels depict two distinct operational regimes depending on whether the harmonic lies within or beyond the secondary plateau. Beyond the cutoff, only a single trajectory branch survives, whose phase slope exhibits excellent consistency with the analytical prediction of Eq.~\eqref{eq:alpha2e_explicit}. Within the plateau, the coexisting short and long trajectories exhibit distinct phase gradients $\alpha_{U_p}$, with also good agreement with Eq.~\eqref{eq:alpha2e_explicit}

The relatively slow linear scaling of the two-electron phase across the secondary plateau makes this emission channel exceptionally suited for attosecond pulse synthesis. Because the attochirp remains remarkably low over extended spectral bandwidths, the generated attosecond pulses suffer minimal phase distortion without requiring complex external chirp compensation. Specifically, near photon energies of $1000~\mathrm{eV}$, the calculated attochirp for long trajectories is $C_L = -0.85~\mathrm{as/eV}$, whereas for short trajectories it yields $C_S = 0.88~\mathrm{as/eV}$. These values are nearly an order of magnitude smaller than typical single-electron values at lower photon energies, enabling seamless compression toward the Fourier-transform limit~\cite{7052318}. Furthermore, the near-symmetric, opposite-sign values of $C_S$ and $C_L$ naturally reflect the distinct continuum propagation dynamics of short and long orbits while maintaining comparable, tight temporal confinement for pulses synthesized from either branch.

To characterize attosecond-pulse synthesis, we coherently superpose the
complex harmonic amplitudes over a selected spectral window according to
$E_{\mathrm{atto}}(t)=\operatorname{Re}\left[\sum_{N\in\mathcal{W}}
e\langle r(N\omega_0)\rangle_{2e}e^{-iN\omega_0t}\right]$, where
$\mathcal{W}$ denotes the chosen harmonic bandwidth. The resulting temporal
profile determines the duration and intrinsic compressibility of the synthesized
attosecond pulses.

\begin{figure}[ht]
\centering

  \begin{subfigure}[b]{\linewidth}
    \centering
    \begin{overpic}[height=4.5cm, keepaspectratio]{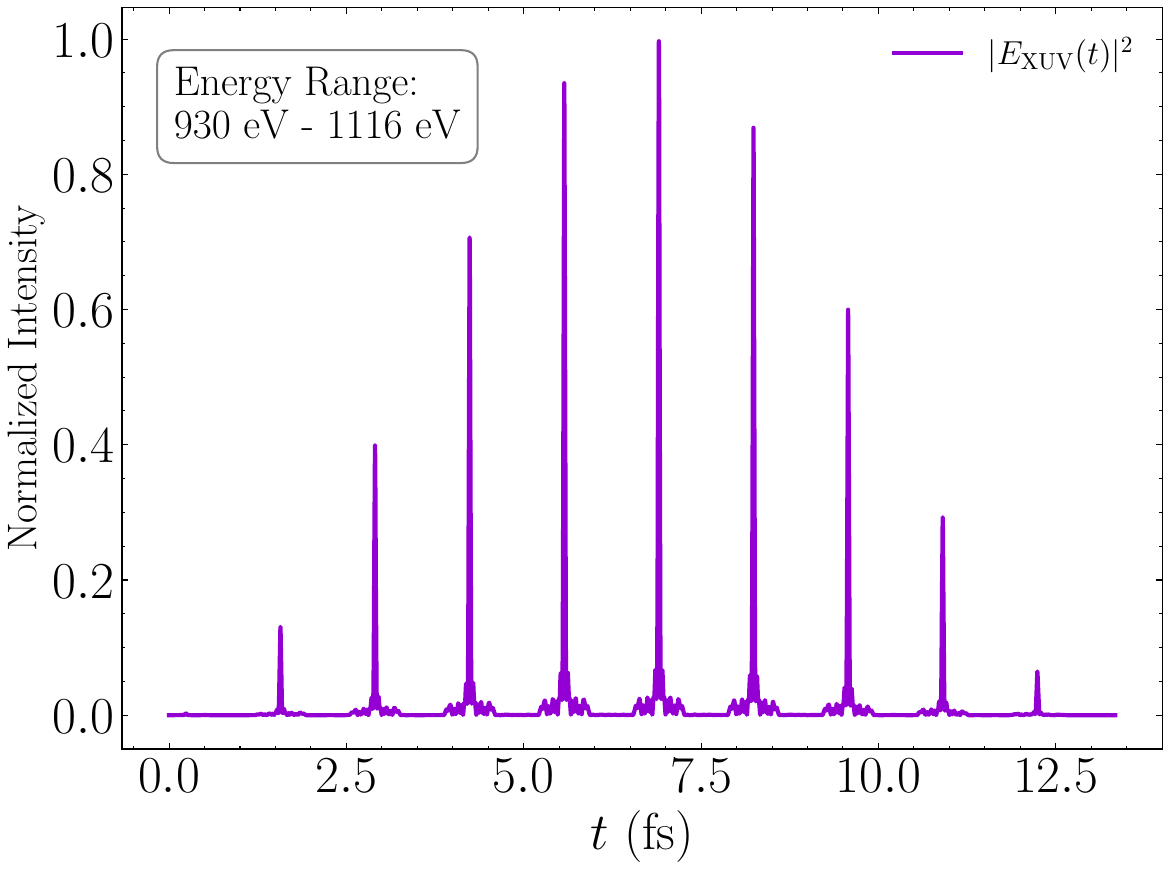}
      \put(15, 50){\textbf{(a)}}
    \end{overpic}
    \phantomcaption
    \label{fig:attopulses_total}
  \end{subfigure}

  \vspace{0.4cm} 

  \begin{subfigure}[b]{\linewidth}
    \centering
    \begin{overpic}[height=4.5cm, keepaspectratio]{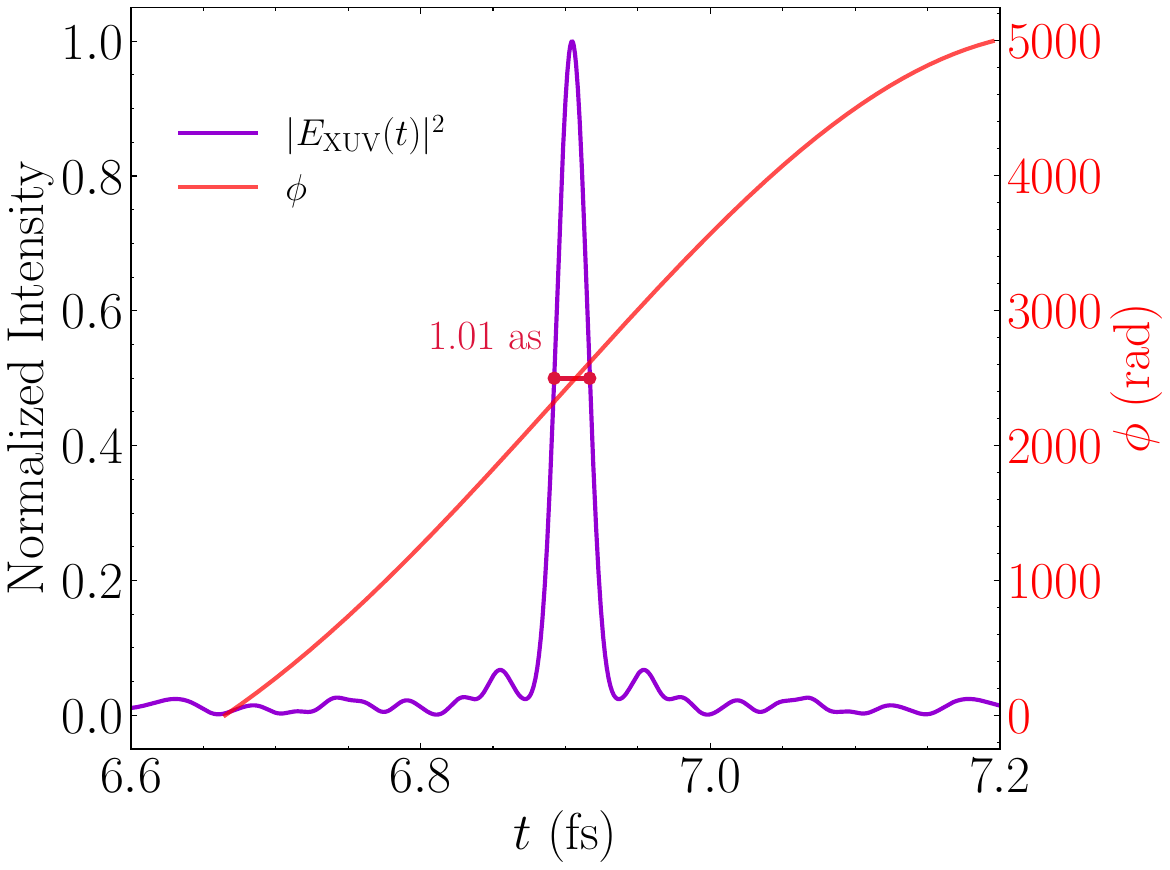}
      \put(15, 50){\textbf{(b)}}
    \end{overpic}
    \phantomcaption
    \label{fig:attopulses_phase}
  \end{subfigure}
  
  \caption{(a) Attosecond pulse train corresponding to the spectrum shown in Fig.3(b) of \cite{McSweeney2026}. (b) Zoom on a single peak displaying the temporal phase obtained from the short trajectories shown in Fig.~2(b).}
  \label{fig:attopulses}
\end{figure}

Fig.~3(a) shows the synthesized attosecond pulse train obtained for the parameters of Ref.~\cite{McSweeney2026}, using a spectral range corresponding to the secondary plateau, between $\sim 930$ eV and $\sim 1100$ eV. Panel~(b) presents a magnified view of an individual burst together with the temporal phase reconstructed from the short-trajectory branch shown in Fig.~\ref{fig:phase_scaling}(b). The broad bandwidth of the correlation-induced secondary plateau supports pulses approaching the sub-attosecond regime. Crucially, bandwidth alone is insufficient: the harmonic components must remain mutually coherent, and their intrinsic attochirp must be sufficiently weak or compensated. In the cutoff region, where the short- and long-trajectory solutions coalesce into a single dominant contribution, this trajectory ambiguity is intrinsically removed. In the plateau, however, two quantum trajectories generally contribute to a given harmonic. In the present single-atom calculations, we select the short-trajectory branch explicitly for simplicity. This should be regarded as a microscopic representation of the trajectory selection that occurs naturally at the macroscopic level through propagation and phase matching. Indeed, Antoine \textit{et al.} showed that propagation through the generating medium can strongly suppress all but one trajectory family, thereby phase-locking the corresponding harmonic components and producing a regular attosecond pulse train~\cite{Antoine1996}. Subsequent propagation calculations demonstrated that the short-return-time contribution can be preferentially phase matched for suitable focusing and gas-jet geometries, owing to its weaker intensity-dependent dipole phase~\cite{Salieres1998}. Thus, although a complete quantitative prediction of the emitted waveform ultimately requires propagation through the macroscopic medium, the short-trajectory synthesis considered here represents a physically accessible phase-matched contribution rather than an arbitrary restriction of the microscopic dynamics. Under such trajectory-selective conditions, double-electron recombination could enable pulse durations of a few hundred zeptoseconds, providing a realistic step toward the zeptosecond frontier.

Experimentally, this prospect requires the coherent enhancement of the weak two-electron plateau, preferential selection of the short trajectories, and phase matching over the extended electron excursion time. Spatial filtering, ionization or polarization gating, and macroscopic propagation control could be used to suppress competing trajectories and isolate a single burst. Residual spectral phase could subsequently be compensated using thin metallic filters, multilayer x-ray optics, or tailored driving fields. A direct experimental reconstruction would be challenging but could be pursued through x-ray streaking or spectral-interferometric techniques adapted to sub-attosecond temporal resolution. Multicycle drivers would naturally produce a zeptosecond-scale pulse train, whereas few-cycle driving and temporal gating would be required to obtain an isolated pulse.

Such sources would provide access to electronic motion on timescales shorter than conventional attosecond pulses, including correlated two-electron rearrangement, ultrafast screening, shake-up and shake-off dynamics, and the earliest stages of inner-shell relaxation. Their high photon energies and broad coherent bandwidth could also enable time-resolved core-level spectroscopy and, provided that sufficient photon flux is achieved, nonlinear x-ray experiments with unprecedented temporal resolution.

In conclusion, we have established a comprehensive two-electron saddle-point framework for HHG dipole emission. The temporal separation between the two ionization events, $\Delta t\approx2\pi/\omega_0$ or $\Delta t\approx\pi/\omega_0$, selects distinct correlated recollision families and thereby controls both the plateau cutoff, approximately $5.55U_p$ or $4.7U_p$, and the phase coefficient $\alpha_{2e}$. Incorporating the complete action phase demonstrates that the extended bandwidth generated by two-electron quantum pathways can be coherently synthesized into sub-attosecond radiation. Electron correlation therefore emerges not merely as a correction to the HHG yield, but as a resource for extending coherent x-ray emission and advancing attosecond pulse synthesis toward the zeptosecond regime.

\begin{acknowledgments}
ICFO-QOT group acknowledges support from:
MCIN/AEI (PGC2018-0910.13039/501100011033,  CEX2019-000910-S/10.13039/501100011033, Plan National STAMEENA PID2022-139099NB, I00, project funded by MCIN/AEI/10.13039/501100011033 and by the “European Union NextGenerationEU/PRTR" (PRTR-C17.I1), FPI); Fundació Cellex; Fundació Mir-Puig; 
Generalitat de Catalunya (European Social Fund FEDER and CERCA program;
Barcelona Supercomputing Center MareNostrum (FI-2023-3-0024); 
EU funding  (EU Horizon 2020 FET-OPEN QU-ATTO, 101168628, HORIZON-CL4-2022-QUANTUM-02-SGA  PASQuanS2.1.

A. M and M. F. C. acknowledge support by the Quantum Science and Technology-National Science and Technology Major Project (Grant No. 2025ZD0301000),  the National Key Research and Development Program of China (Grant No. 2023YFA1407100), the Guangdong Province Science and Technology Major Project (Future functional materials under extreme conditions - 2021B0301030005) and the National Natural Science Foundation of China (Grant No. 12574092). 

P.T. at FORTH acknowledges the European Union’s HORIZON-MSCA-2023-DN-01 project QU-ATTO under the Marie Skłodowska-Curie grant agreement No 101168628 and ELI–ALPS. The ELI-ALPS project (GINOP-2.3.6-15-2015-00001) is supported by the EU and co-financed by the European Regional Development Fund. 

\end{acknowledgments}

\bibliographystyle{apsrev4-2}
\bibliography{ref}

\end{document}